\documentclass[twocolumn,superscriptaddress,prl]{revtex4}
\usepackage{graphicx}
\usepackage{dcolumn}
\usepackage{amsmath}
\usepackage{amssymb}
\usepackage[utf8]{inputenc}
\usepackage[english]{babel}

\newcommand{\Tr}{\,\mathrm{Tr}}

\begin{document}

\title{The integrated density of states of the random graph {L}aplacian}
\author{T. Aspelmeier}
\affiliation{Max Planck Institut für Dynamik und Selbstorganisation,
Bunsenstr.\ 10, 37073 Göttingen, Germany}
\affiliation{Scivis GmbH, Bertha-von-Suttner-Str.\ 5, 37085 Göttingen, Germany}
\author{A. Zippelius}
\affiliation{Max Planck Institut für Dynamik und Selbstorganisation,
Bunsenstr.\ 10, 37073 Göttingen, Germany}
\affiliation{Institut für Theoretische Physik, Georg-August
  Universität Göttingen, Friedrich-Hund-Platz 1, 37077 Göttingen, Germany}


\begin{abstract}
  We analyse the density of states of the random graph Laplacian in
  the percolating regime. A symmetry argument and knowledge of the
  density of states in the nonpercolating regime allows us to isolate
  the density of states of the percolating cluster (DSPC) alone,
  thereby eliminating trivially localised states due to finite
  subgraphs. We derive a nonlinear integral equation for the
  integrated DSPC and solve it with a population dynamics
  algorithm. We discuss the possible existence of a mobility edge and
  give strong evidence for the existence of discrete eigenvalues in the whole range of the
  spectrum.
\end{abstract}

\maketitle

The eigenvalue spectrum of sparse random matrices is a fascinating and
largely unsolved problem with widespread applications ranging from
transport in disordered systems, graph theory and optimization
problems to nuclear physics and QCD \cite{Mehta,Guhr98}. In this
paper we consider a prototype of such a random matrix, the Laplace
operator on a mean-field random graph with $N$ nodes, i.e.\ a graph
where a link between two arbitrary sites is either present with
probability $p=2c/N$ or not present with probability $1-p$. The
constant $2c$ is the mean connectivity of the graph. The Laplace
operator on this graph is a matrix $\Gamma_{ij}$ where for $i\ne j$
$\Gamma_{ij} =-1$ if the nodes $i$ and $j$ of the graph are connected
and $\Gamma_{ij} = 0$ otherwise, while on the diagonal $\Gamma_{ii} =
-\sum_{j\ne i} \Gamma_{ij}$. The entries on the diagonal are thus
correlated to the random entries outside the diagonal.

Even though the computation of the density of states for a mean-field
random graph has been reduced to an integral equation
\cite{Bray:1988,Fyodorov:1991,Khorunzhy:1994}, a complete solution is
still missing.  In the limit of infinite coordination, $c\to\infty$,
Wigner`s semi-circle law is recovered. For any finite $c$, Lifshitz
tails were shown to exist in the integrated density of
states~\cite{Khorunzhy:2006}. Beyond these asymptotic results a number
of approximations have been used to compute the spectrum
approximately, such as effective medium theory, single defect
approximation, moment expansions and numerical
diagonalisation~\cite{Biroli:1999,Bauer:2001,Kuehn:2008}.
% Biroli and Monasson identified a mobility edge with
%help of the participation ratio and presented a heuristic argument to
%associate localised states with sites of unusually large or small
%connectivity. K\"uhn has used a population dynamics algorithm on the
%density of eigenvalues and confirms the existence of a mobility edge.

In this paper we show first that the density of states of the
percolating cluster (DSPC) can be isolated using a symmetry
argument and our knowledge of the density of states below the
percolation threshold. Second, we derive an integral equation for the
integrated density of states which can be solved reliably with a
population algorithm. The numerical solution reveals jumps in the
integrated DSPC in the whole range of the spectrum, calling in
question the existence of a mobility edge.

{\bf Model and symmetry}: The model shows a percolation transition at
$c_{crit}=1/2$. Below this concentration there is no macroscopic
cluster and almost all finite clusters are trees. The average number
of tree clusters $T_n$ with $n$ nodes is given in the macroscopic
limit by \cite{ErRe60}
\begin{equation}\label{trees}
\lim_{N\to\infty} \frac{T_n(2c)}{N}=  \tau_{n}(2c)=
\frac{n^{n-2}(2c\,{\rm e}^{-2c})^n}{2c\,n!}. 
\end{equation}
In particular the total number of clusters per particle is
$\tau_{\text{tot}}(2c)=1-c$.  For $c<1/2$, the spectrum consists of a 
very complicated,
but countable set of $\delta$-peaks which can be calculated iteratively
\cite{Broderix:2001}. Above the percolation threshold $c>1/2$ a
percolating cluster coexists with many finite clusters, which are also
trees.
% Furthermore the number of trees above and below the percolation
%threshold are simply related.
The fraction of sites in the macroscopic
cluster, $Q(c)$, is the solution of
$1-Q(c)=\exp(-2cQ(c))$. Alternatively we rewrite $Q(c)=1-\frac{x(c)}{2c}$
and obtain $x(c)$ as the solution of
\begin{equation}
\label{symx}
x(c)e^{-x(c)}=2ce^{-2c}.
\end{equation}
This equation has two solutions, a trivial one with $x(c)=2c$ and a
nontrivial one with $x(c)=2c^*$ such that $c^*>\frac
12$ if $c<\frac 12$ and vica versa. This nontrivial solution allows one 
to establish a symmetry for the
number of trees above and below the percolation threshold. % According
%to Eq.(\ref{trees}) this number is given by
%\begin{equation}
%\tau_n(x(c))=\frac{n^{n-2}(x(c)\,{\rm e}^{-x(c)})^n}{x(c)\,n!}. 
%\end{equation}
Using  Eq.~(\ref{symx}), we can rewrite Eq.~(\ref{trees})
according to
\begin{equation}
\label{symmetry}
\tau_n(x(c))=\frac{2c}{x(c)}\tau_n(2c) \quad \text{or} \quad \tau_n(2c^*)=\frac{c}{c^*}\tau_n(2c).
\end{equation}
Hence the number of trees {\it  above} the percolation threshold is simply
related to the number of trees {\it below} the percolation threshold.

{\bf Density of states}
 This relation allows us to compute the
density of states of the percolating cluster {\it alone}, which is the
quantity of primary interest. It is known that the density of states,
$D(\Omega)$, of the infinite cluster contains at least some $\delta$
peaks, so that a population dynamics algorithm \cite{Kuehn:2008}
cannot be applied.  In this paper we instead compute the
\textit{integrated} density of states,
$\Delta(\Omega)=\int_0^{\Omega} d\Omega^{\prime}
D(\Omega^{\prime})$, which according to measure theory \cite{Reed:1980},
may be decomposed into a singular part and an
absolutely continuous part. The absolutely contiuous part may itself
consist of two contributions, eigenvalues stemming from localised
eigenvectors which happen to lie dense and each have vanishing weight
in the thermodynamic limit, and the eigenvalues stemming from
nonlocalised eigenvectors. We may thus write
\begin{align}
\Delta(\Omega) &= \Delta_{\text{loc,disc}}(\Omega) +
\Delta_{\text{loc,cont}}(\Omega) + \Delta_{\text{nonloc}}(\Omega).
\end{align}
Often in random matrix problems there is a mobility edge, i.e.\ a value
$\Omega_0$ such that all eigenvectors corresponding to eigenvalues
$\Omega<\Omega_0$ are localised and all eigenvectors corresponding to
$\Omega>\Omega_0$ are nonlocalised (or vice versa). We will show here that such
a sharp edge does not seem to exist in our problem as we find discrete
eigenvalues even in the region where nonlocalised eigenvectors lie.
This shows that a naive approach using the inverse participation ratio in order
to find the onset of nonlocalised eigenvectors will not work since the
localised eigenvectors do not disappear where the nonlocalised ones start, so
the inverse participation ration will not drop down to $0$ as it would at a
true mobility edge.

The density of eigenvalues, $\{\Omega_i\}_{i=1}^N$, of the
Laplacian matrix $\Gamma$ is defined by
\begin{equation}
\label{eq:densitydef}
D(\Omega,c)=
  \lim_{\small{N\to\infty}}\overline{\frac{1}{N}\sum_{i=1}^{N}
\delta(\Omega-\Omega_i)}
 = \lim_{\small{N\to\infty}}\overline{\frac{1}{N}\Tr\delta(\Omega-\Gamma)}
\end{equation}
Here $\overline{\,\cdot\,}$ denotes the
average over all realizations of connectivity for a given $c$. 
To compute the density of eigenvalues we introduce the resolvent
\begin{equation}
\label{resolvent}
G(\Omega,c)=\lim_{N\to\infty}\frac{1}{N}\overline{\Tr 
\frac{1}{\Gamma-\Omega}}
\end{equation}
for complex argument $\Omega=\gamma + i\epsilon, \epsilon>0$. 
In the limit $\epsilon\to 0$, we recover the spectrum from the
imaginary part of the resolvent according to
\begin{equation}
D(\Omega,c) = 
  \frac{1}{\pi}\lim_{\epsilon\downarrow 0} \Im G(\Omega+i\epsilon,c).
\end{equation}

In the percolating regime $c\geq \frac 12$ the macroscopic cluster
coexists with many finite ones. In order to study localised states of
the macroscopic cluster it is essential to isolate the density of
states of the macroscopic cluster only. We use Eqs.~(\ref{symx}) and \eqref{symmetry}
to decompose the resolvent into two
contributions, one from the percolating cluster and one from the
finite clusters:
\begin{equation}
G(\Omega,c)=G^{\text{perc}}(\Omega,c)+\frac{c^*}{c} G(\Omega,c^*)
\end{equation}
%A similar equation holds for the density of states of the percolating
%cluster. 
The above relation allows us to compute the density of
states of the percolating cluster alone from the full density of states
above, $G(\Omega,c)$, and the density of states below the percolation threshold,
$G(\Omega,c^*)$ for $c^*<\frac 12$, which is known
\cite{Broderix:2001}. 

The average over all realizations of connectivity is performed with
the replica trick, resulting in a
nonlinear integral equation for $g_{c,\Omega}(\rho)$ (cf.\ Eqs.~(16) and (17)
in Ref.~\cite{Bray:1988})
\begin{widetext}
\begin{equation}
\label{Eq35}
  g_{c,\Omega}(\rho)   =  2c \,  \exp\!\left\{-\frac{i\rho^2}{2}\right\} 
 +
  2ic\, e^{-2c} \int_0^\infty d x\,
  \rho \,\text{I}_1(i\rho x)\,
  \exp\!\left\{
    -\frac{i}2\,(\rho^2+x^2)
    +\frac{i \Omega}2\,x^2 +g_{c,\Omega}(x) 
  \right\} 
\end{equation}
\end{widetext}
with $g_{c,\Omega}(0)=2c$. Here $\text{I}_\nu(z)$ are the modified
Bessel functions of the first kind.  The solution of Eq.~(\ref{Eq35})
yields the resolvent~\cite{Bray:1988}
%%%%%%%%%%%%%%%%%%%%%%%%%%%%%%%%%%%%%%%%%%%%%%%%%%%%%%%%%%%%%%%%%%%%%%%%%%%%%
%  Timo, bitte check die Vorzeichen in der folgenden Gleichung. Meiner
%  Meinung nach waren sie in der vorangehenden Version falsch.
%%%%%%%%%%%%%%%%%%%%%%%%%%%%%%%%%%%%%%%%%%%%%%%%%%%%%%%%%%%%%%%%%%%%%%%%%%%%%%
\begin{equation} \label{Eq32}
  G(\Omega,c)  =- 1 +
  \frac{i}{2c}\,\int_0^\infty d\rho\,\rho\,g_{c,\Omega}(\rho) 
\end{equation}

{\bf Nonpercolating regime}: The analytical
solution of Eq.~\eqref{Eq35} for $c<1/2$ was given in
\cite{Broderix:2001}. Briefly, it is 
\begin{align}
g_{c,\Omega}(\rho) &=
2c\sum_k a_k \exp(-\frac i2 z_k \rho^2) = 2c\int_{-\infty}^\infty
da_{c,\Omega}(\lambda) e^{-\frac i2 \lambda\rho^2}
\label{laplace}
\end{align}
with coefficients $a_k$ and $z_k$ to be defined below.
The sum can also be expressed as a Riemann-Stieltjes
integral with weight function $a_{c,\Omega}(\lambda) = \sum_k a_k
\theta(\lambda-z_k)$ ($\theta(\cdots)$ is the Heaviside function). This
formulation will be useful later.

The coefficients $a_k$ and $z_k$ can be grouped in infinitely many ``classes.''
The coefficients of class $n+1$ are given recursively by the relations
\begin{align}
a^{(n+1)}_{(l_k)} = e^{-2c}\prod_{k} \frac{(2ca_k^{(n)})^{l_k}}{l_k!}
\label{eq:classa}
\\
z^{(n+1)}_{(l_k)} = \frac{\Omega - \sum_{k} l_k z_k^{(n)}}{\Omega - 1 -
\sum_{k} l_k z_k^{(n)}}
\label{eq:classz}
\end{align}
(the upper index denotes the class). Class $0$ contains only one element, namely
$a^{(0)}_0 = e^{-2c}$ and $z^{(0)}_0 = \frac{\Omega}{\Omega-1}$. The index on the
left hand side is a finite sequence $(l_k)$ of nonnegative integers. In order to
proceed to the next stage of the recursion, $(l_k)$ must be mapped to a
natural number $m$ since for the calculation of, say, $z_{(l_k)}^{(n+1)}$ it is necessary to 
access the coefficients $z_m^{(n)}$ of the previous iteration. 
Such a mapping is possible because the set of sequences $\{(l_k)\}$
is countably infinite. It was shown in \cite{Broderix:2001} that the correct way to do the
mapping is to choose $m=\sum_k l_k M^k$ and to let $M$ (formally) tend to infinity at an
appropriate point.  Note that 
class $n+1$ also contains all coefficients from class $n$.
See \cite{Broderix:2001} for details.  

These classes give us an infinite hierarchy of coefficients, each
recursion step adding infinitely many coefficients to the previous
ones. Note that the coefficients in class $n$ constructed in this way
are \textit{not} an approximation but constitute (part of) the exact
solution of Eq.~\eqref{Eq35}.  Only the total weight of coefficients
$\sum_k a_k$ falls short of $1$ when stopping the recursion at a
finite $n$. This solution of Eq.~\eqref{Eq35} leads to a density of
states consisting of $\delta$ peaks which are located at those
$\Omega$ where $z_k=0$.

{\bf Percolating regime}: In complete analogy to the resolvent, we can
decompose $g_{c,\Omega}(\rho)= g_{c,\Omega}^{\mathrm{perc}}(\rho) +
g_{c^*,\Omega}(\rho)$ into a part $g_{c,\Omega}^{\mathrm{perc}}(\rho)$
pertaining to the infinite cluster and a part which is equal to the
(known) solution $g_{c^*,\Omega}(\rho)$ of the integral equation at
$c^*$.  This decomposition has the advantage that
we can directly obtain the density of states of the infinite cluster
by deriving an equation for $g_{c,\Omega}^{\mathrm{perc}}(\rho)$,
which in analogy to Eq.~\eqref{laplace} can be represented as
\begin{align}
g_{c,\Omega}^{\mathrm{perc}}(\rho) = (2c - 2c^*)\int_{-\infty}^\infty
db_{c,\Omega}(\lambda)\, e^{-\frac i2 \lambda \rho^2}.
\label{defa}
\end{align}
The function $b_{c,\Omega}(\lambda)$ is normalized such that
$\int_{-\infty}^\infty db_{c,\Omega}(\lambda) = 1$ and the
``initial condition'' $g_{c,\Omega}(0) = 2c$ is being taken care of by the
prefactor $2c-2c^*$. This ansatz is plugged into the integral Eq.~(\ref{Eq35})
to yield

\begin{widetext}
\begin{equation}
b_{c,\Omega}(\lambda) =
2c^* \sum_{n=0}^\infty a_{n}
\sum_{M=1}^\infty \frac{(2c-2c^*)^{M-1}}{M!}\int_{-\infty}^\infty
db_{c,\Omega}(\lambda_1) \cdots \int_{-\infty}^\infty 
db_{c,\Omega}(\lambda_M) \times 
\theta\left( \lambda - \left[1-\frac{1}{\frac{1}{1-z_{n}} 
+  \sum_{i=1}^M \lambda_i}\right] \right).
\label{eq:percolatingequation}
\end{equation}
\end{widetext}
This equation can be solved numerically by running a population
dynamics algorithm for the coefficients $z_k$ at $c^*$
\textit{below} the critical point in parallel to a population dynamics
for a $\lambda$-population at $c$ \textit{above} the critical point,
for which the $z_k$ and their weights $a_k$ are needed as input.

%The contribution to the density of states from the percolating cluster in terms
%of $b_{c,\Omega}(\lambda)$ is
%\begin{align}
%D^{\mathrm{perc}}(\Omega, c) &= \frac{1}{\pi}\Im \int_0^\infty d\rho\,\rho
%\int_{-\infty}^\infty db_{c,\Omega}(\lambda) e^{-\frac i2
%\lambda\rho^2}\nonumber\\
% &= \int_{-\infty}^\infty
%db_{c,\Omega}(\lambda)\delta(\lambda)
%\end{align}
%Note that the relative weight of the infinite cluster has been divided
%out, therefore with knowledge of $b_{c,\Omega}(\lambda)$ we can obtain
%$D^{c,\text{perc}}$ for any $c\ge \frac 12$, even directly at the
%critical bond concentration $c=\frac 12$ although the relative weight
%of the infinite cluster is $0$ there.

%The above density of states would be very simple if $b_{c,\Omega}(\lambda)$ was
%differentiable as it would then be equal to $(b_{c,\Omega})'(0)$. It is however
%known that the density of states of the percolating cluster contains $\delta$
%peaks also for $c>1/2$. Hence we face the same problem as below the percolation
%threshold that the position and weight of the $\delta$ peaks can not reliably be
%obtained from a population dynamics algorithm.
Given $a_{c^*,\Omega}(\lambda)$ and $b_{c,\Omega}(\lambda)$ the
integrated DSPC can be computed according to (see \cite{supplement}):
\begin{widetext}
\begin{align}
2 \Delta^{\mathrm{perc}}(\Omega, c) &=  \left( 1 +
\int_{-\infty}^\infty db_{c,\Omega}(\lambda)\left(
\mathrm{sgn}(\frac{\lambda}{\lambda-1}) + c\,
\mathrm{sgn}(\lambda-1)\right) - (c-c^*)
\int_{-\infty}^\infty db_{c,\Omega}(\lambda)
db_{c,\Omega}(\lambda')
\mathrm{sgn}(\frac{\lambda}{\lambda-1}-\lambda') 
\right. \nonumber\\
&\quad \left. + c^*\int
da_{c^*,\Omega}(\lambda)\,\mathrm{sgn}(\frac{1}{\lambda-1}) -  
c^*\int(da_{c^*,\Omega}(\lambda)db_{c,\Omega}(\lambda') +
db_{c,\Omega}(\lambda)da_{c^*,\Omega}(\lambda'))
\mathrm{sgn}(\frac{\lambda}{\lambda-1}-\lambda')
\right).
\label{eq:integrateddosb}
\end{align}
\end{widetext}

{\bf Discussion}: We have developed a systematic approach to compute the
integrated DSPC. We stress that the (nonintegrated) DSPC could not be reliably
obtained from a population dynamics algorithm. Naively it is given in terms of
$b_{c,\Omega}(\lambda)$ by
\begin{align}
D^{\mathrm{perc}}(\Omega, c) &= \frac{1}{\pi}\Im \int_0^\infty d\rho\,\rho
\int_{-\infty}^\infty db_{c,\Omega}(\lambda) e^{-\frac i2
\lambda\rho^2}\nonumber\\
 &= \int_{-\infty}^\infty
db_{c,\Omega}(\lambda)\delta(\lambda)
\end{align}
%Note that the relative weight of the infinite cluster has been divided
%out, therefore with knowledge of $b_{c,\Omega}(\lambda)$ we can obtain
%$D^{c,\text{perc}}$ for any $c\ge \frac 12$, even directly at the
%critical bond concentration $c=\frac 12$ although the relative weight
%of the infinite cluster is $0$ there.
The above density of states would be very simple if
$b_{c,\Omega}(\lambda)$ was differentiable with respect to $\lambda$,
as it would then be equal to $(b_{c,\Omega})'(0)$. It is however known
that the density of states of the percolating cluster contains
$\delta$ peaks also for $c>1/2$. Hence the position and weight of the
$\delta$ peaks can not reliably be obtained from a population dynamics
algorithm -- a problem already encountered in the nonpercolating
regime. It is thus essential to obtain the \textit{integrated} density of
states directly from population dynamics without going to the density of
states first and integrating, and this is what 
Eqs.~\eqref{eq:percolatingequation} and \eqref{eq:integrateddosb} 
provide.

\begin{figure}
\includegraphics[width=\linewidth]{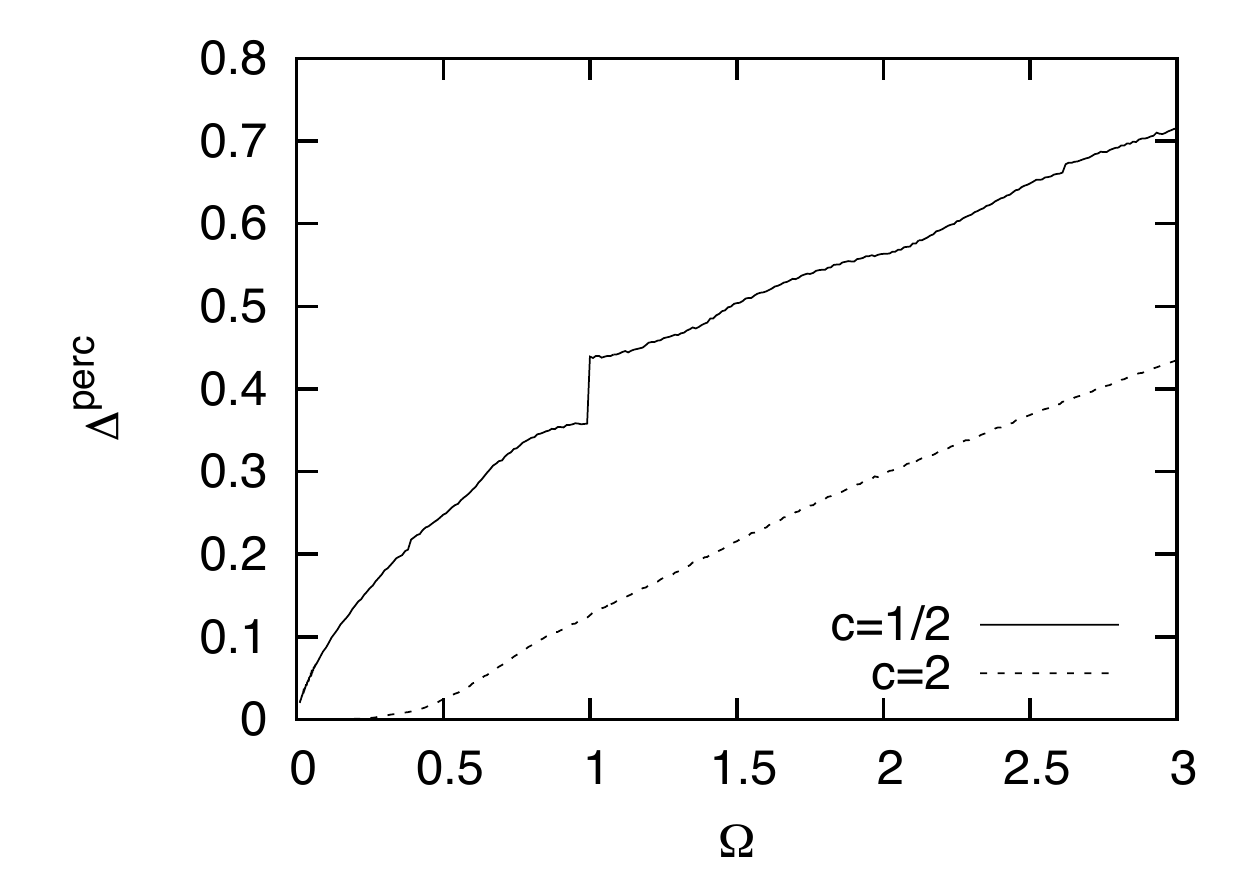}
\includegraphics[width=\linewidth]{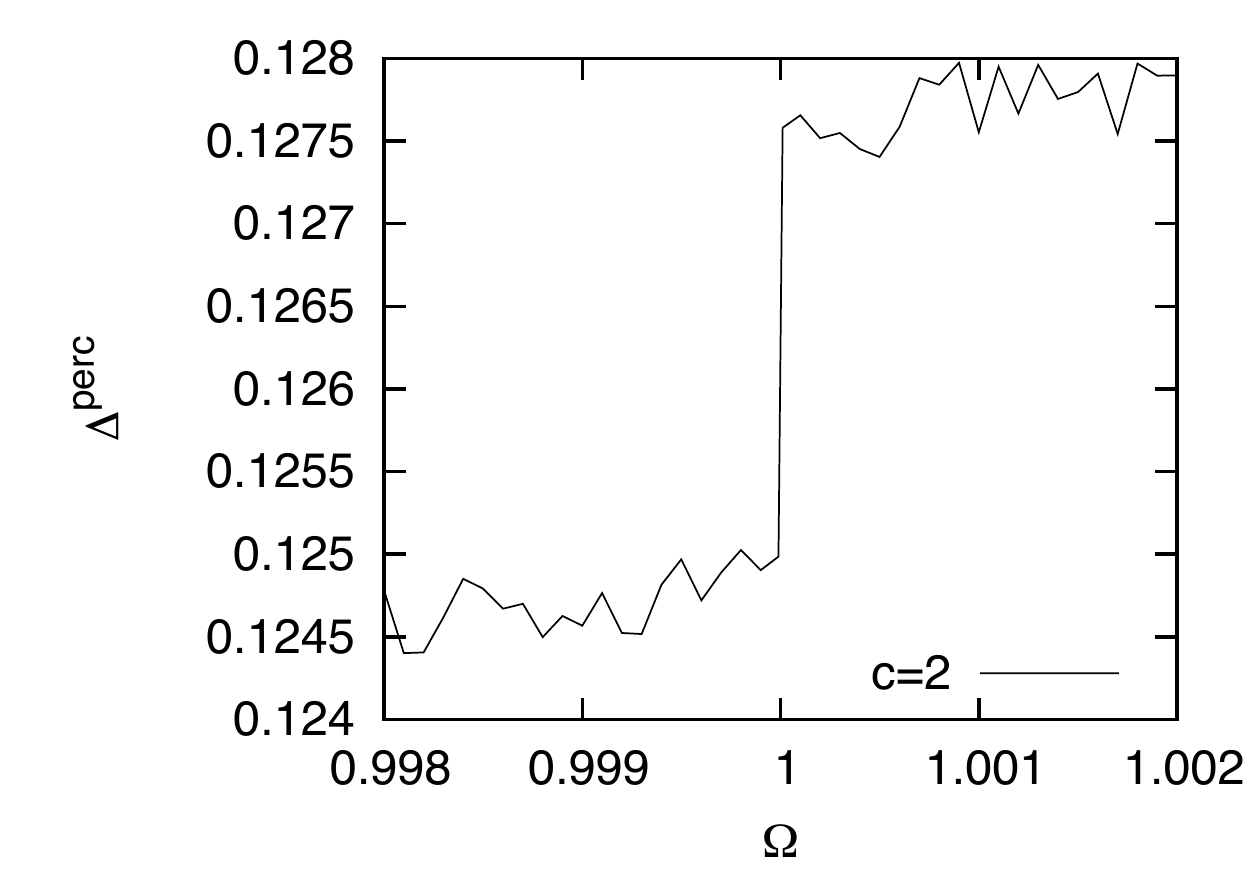}
\caption{The integrated density of states at $c=\frac 12$ and $c=2$. The bottom figure shows
an enlargement of the curve for $c=2$ around $\Omega=1$.}
\label{Fig1}
\end{figure}

%\begin{figure}
%\includegraphics[width=\linewidth]{ids4}
%\caption{Enlargement of the integrated density of states at
%$c=2$.}
%\label{Fig2}
%\end{figure}

%In order to demonstrate the method, we show in Fig.~\ref{Fig1} the integrated
%density of states directly at the critical point $c=\frac 12$ and deep inside
%the percolating regime at $c=2$. Some jumps are clearly observed for $c=\frac
%12$, and they are located at the predicted positions, for instance at $\Omega=1$
%where the coefficient $z=\frac{\Omega}{\Omega-1}=\infty$, or at
%$\Omega=\frac{3\pm\sqrt{5}}{2}$ where $z=\frac{\Omega -
%\frac{\Omega}{\Omega-1}}{\Omega-1 -
%\frac{\Omega}{\Omega-1}}=\infty$.

%For $c=2$ the integrated density of states in Fig.~\ref{Fig1} looks smooth. This
%is, however, not the case as the close-up in Fig.~\ref{Fig2} reveals. Here, too,
%a jump at $\Omega=1$ is evident. Other jumps are not discernible since they are
%so small that they disappear in the noise (since the population dynamics
%algorithm is a stochastic method, a certain amount of noise is inevitable,
%depending on the size of the population).

The final equation for the integrated DSPC,
Eq.~\eqref{eq:integrateddosb}, reveals much about the eigenvalues of
the percolating cluster. We can, for example, track down the origin of
the $\delta$ peaks in the density of states of the percolating
cluster. Suppose for the moment that $b_{c,\Omega}(\lambda)$ is
continuous as a function of $\Omega$. Then the integrals over
$db_{c,\Omega}(\lambda)$ certainly do not generate any jumps in
$\Delta^{\mathrm{perc}}(\Omega,c)$ due to the continuity in
$\Omega$. However, we know that $a_{c^*,\Omega}(\lambda)=\sum_n a_n
\theta(\lambda-z_n)$ is discontinuous as a function of $\Omega$ since
the $z_n$ depend on $\Omega$, and inspection of
Eq.~\eqref{eq:integrateddosb} shows that this leads to a jump in
$\Delta^{\mathrm{perc}}(\Omega)$ if the location $z_n$ of a jump moves
from $-\infty$ to $+\infty$ when $\Omega$ is increased infinitesimally
(it does not lead to a jump if $z_n$ moves across $0$ or $1$, as could
be suspected at first sight, since the various contributions cancel in
these cases).  Eq.~\eqref{eq:classz} shows that $z_n$ can and does
indeed pass $\infty$ as $\Omega$ is varied. While the peaks of the
density of states of the \textit{finite} clusters are located at those
$\Omega$ for which $z_n=0$, the peaks for the percolating cluster are
located where $z_n=\infty$. Since the zeros and the poles of $z_n$
necessarily alternate when $\Omega$ is varied, it follows that the
peaks in the percolating cluster lie dense if the peaks in the finite
clusters lie dense.  In this sense there is no mobility edge in the
percolating cluster since isolated and thus localized eigenvalues
exist throughout the whole range of $0\le\Omega<\infty$.

Unfortunately, this argument only strictly holds if $b_{c,\Omega}(\lambda)$ is
indeed continuous in $\Omega$. If it is not, cancellations might occur which
could reduce (or even remove) the peaks from the percolating cluster. It is shown
in \cite{supplement} that indeed $b_{c,\Omega}(\lambda)$ is not continuous but the
argument presented there also shows that a complete removal of peaks would seem
an extremely fortuitous cancellation. In order to check these results we have 
performed population dynamics simulations of the integrated DSPC according 
to Eq.~\eqref{eq:percolatingequation}.
Fig.~\ref{Fig1} shows the integrated DSPC directly at the critical point $c=\frac
12$ and deep inside the percolating regime at $c=2$. Some jumps are clearly
observed for $c=\frac 12$, and they are located at the predicted positions. The
most prominent ones occur at $\Omega=1$ where the coefficient
$z=\frac{\Omega}{\Omega-1}=\infty$, or at $\Omega=\frac{3\pm\sqrt{5}}{2}$ where
$z=\frac{\Omega - \frac{\Omega}{\Omega-1}}{\Omega-1 -
\frac{\Omega}{\Omega-1}}=\infty$. For $c=2$ the integrated density of states in
Fig.~\ref{Fig1} looks smooth. This is, however, not the case as the close-up in
the bottom part of Fig.~\ref{Fig1} reveals. The jump at $\Omega=1$ which is very 
pronounced at $c=\frac 12$ is also present at $c=2$.

\begin{acknowledgments}
Our work is based on ideas by Kurt Broderix
who died on May 12, 2000. We thank Peter Müller and Reimer Kühn for
valuable discussions.
\end{acknowledgments}

%\bibliographystyle{unsrt}

%\bibliography{LiteraturDB,cond-mat}

\end{document}